\documentclass{amsart}

\usepackage[english]{babel}
\usepackage[latin1]{inputenc}
\usepackage[dvips,final]{graphics}
\usepackage{amsmath,amsfonts,amssymb,amsthm,amscd,array,stmaryrd,mathrsfs}
\usepackage{pstricks}
 \usepackage[all]{xy}
\usepackage{textcomp}
 \usepackage[final]{epsfig}
\theoremstyle{plain}

\newtheorem*{Thm}{Theorem}

\theoremstyle{definition}
\newtheorem{ques}{Question}
\newtheorem*{defn}{Definition}
\newtheorem*{rem}{Remark}
\newtheorem*{ex}{Example}

\newcommand{\R}{\mathbb{R}}
\newcommand{\Z}{\mathbb{Z}}

\newcommand{\E}{\mathcal{E}}
\newcommand{\Fc}{\mathcal{F}}
\newcommand{\Kc}{\mathcal{K}}

\newcommand{\M}{\mathcal{M}} 
\newcommand{\Pc}{\mathcal{P}}

\newcommand{\Ic}{\mathcal{I}}

\newcommand{\ga}{\mathfrak{a}}

\newcommand{\gf}{\mathfrak{g}}

\newcommand{\half}{\frac{1}{2}}

\newcommand{\e}{\varepsilon}

\newcommand{\asl}{\mathrm{asl}}
\newcommand{\osp}{\mathrm{ \bf osp}(1,2)}
\newcommand{\Der}{\textup{\bf Der}}

\newcommand{\ddp}{\frac {\partial}{\partial p}} 
\newcommand{\ddq}{\frac {\partial}{\partial q}} 
\newcommand{\ddt}{\frac {\partial}{\partial \tau}}

\def\e{\varepsilon}

\def\g{\gamma}
\def\L{\Lambda}
\def\om{\omega}

\def\vp{\varphi}

\def\l{\lambda}

\begin{document}

\title{A short survey of Lie antialgebras}

\author{S{\'e}verine Leidwanger
and  Sophie Morier-Genoud}

\address{
S{\'e}verine Leidwanger,
Institut de Math\'ematiques de Jussieu - Paris Rive Gauche, UMR 7586 
Projet Groupes, repr\'esentations et g\'eom\'etrie, 
B\^atiment Sophie Germain,
Case 7012 
75205 Paris Cedex 13, France}

\email{leidwang@math.jussieu.fr}

\address{
Sophie Morier-Genoud,
Institut Math\'e\-ma\-tiques de Jussieu - Paris Rive Gauche,
UMR 7586,
Universit\'e Pierre et Marie Curie Paris 6,
4 place Jussieu, case 247,
75252 Paris Cedex 05, France}

\email{sophiemg@math.jussieu.fr}

\date{}


\begin{abstract}
This article is the writing notes of a talk on Lie Antialgebras given
by the second author 
at the conference \textit{3Quantum: Algebra Geometry Information} that 
held in Tallinn in July 2012. The aim of this note is to give a brief survey of the existing theory 
of Lie antialgebras
and to suggest open questions.

\end{abstract}

\maketitle

\section{Introduction}

The notion of \textit{Lie antialgebra} is quite recent and due to Valentin Ovsienko. 
Since the founding work \cite{Ovs}, various directions of study on these algebras have been, and are still, investigated
\cite{MG}, \cite{LMG}, \cite{LMG2}, \cite{LO}, \cite{Kre}.

The aim of this note is to give a brief survey of the existing theory and to suggest open questions.

What is a Lie antialgebra?
The name is quite surprising as Lie antialgebras are neither Lie algebras nor antialgebras.
The name was chosen to suggest a ``dark face'' of Lie algebras, a twin brother living in the shadow.
The prefix \textit{anti} also refers to parity inversion and changes of signs occuring in this theory in comparison to the theory of Lie algebras.

Ovsienko likes to illustrate the general idea behind Lie antialgebras with the following diagram.

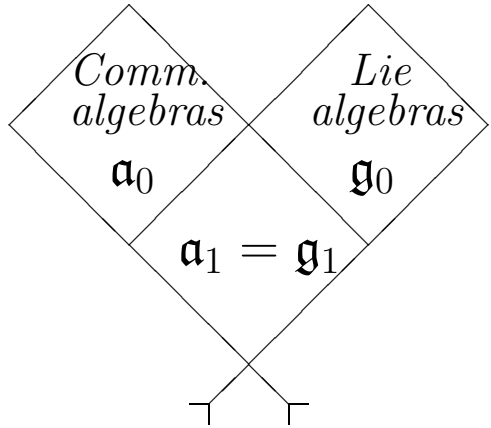
\setlength{\unitlength}{2200sp}%
\begingroup\makeatletter\ifx\SetFigFont\undefined%
\gdef\SetFigFont#1#2#3#4#5{%
  \reset@font\fontsize{#1}{#2pt}%
  \fontfamily{#3}\fontseries{#4}\fontshape{#5}%
  \selectfont}%
\fi\endgroup%
\begin{center}
\begin{picture}(5424,4749)(1789,-6598)
\put(5626,-2761){\makebox(0,0)[lb]{\smash{{\SetFigFont{16}{24.0}{\rmdefault}{\mddefault}{\itdefault}{\color[rgb]{0,0,0}Lie}%
}}}}
\thinlines
{\color[rgb]{0,0,0}\put(4951,-6361){\line( 1,-1){225}}
}%
{\color[rgb]{0,0,0}\put(4951,-6361){\line( 0,-1){225}}
}%
{\color[rgb]{0,0,0}\put(4051,-6361){\line( 0,-1){225}}
}%
{\color[rgb]{0,0,0}\put(4051,-6361){\line(-1,-1){225}}
}%
{\color[rgb]{0,0,0}\put(4051,-6361){\line(-1, 0){225}}
}%
{\color[rgb]{0,0,0}\put(4501,-5911){\line( 1,-1){450}}
}%
{\color[rgb]{0,0,0}\put(4501,-5911){\line(-1,-1){450}}
}%
{\color[rgb]{0,0,0}\put(3151,-4561){\line( 1,-1){1350}}
\put(4501,-5911){\line( 1, 1){1350}}
}%
{\color[rgb]{0,0,0}\put(4501,-3211){\line( 1, 1){1350}}
\put(5851,-1861){\line( 1,-1){1350}}
\put(7201,-3211){\line(-1,-1){1350}}
\put(5851,-4561){\line(-1, 1){1350}}
}%
{\color[rgb]{0,0,0}\put(3151,-1861){\line(-1,-1){1350}}
\put(1801,-3211){\line( 1,-1){1350}}
\put(3151,-4561){\line( 1, 1){1350}}
\put(4501,-3211){\line(-1, 1){1350}}
}%
\put(3700,-4786){\makebox(0,0)[lb]{\smash{{\SetFigFont{20}{24.0}{\rmdefault}{\mddefault}{\itdefault}{\color[rgb]{0,0,0}$\mathfrak{a}_1=\mathfrak{g}_1$}%
}}}}
\put(2926,-3886){\makebox(0,0)[lb]{\smash{{\SetFigFont{20}{24.0}{\rmdefault}{\mddefault}{\itdefault}{\color[rgb]{0,0,0}$\mathfrak{a}_0$}%
}}}}
\put(5626,-3886){\makebox(0,0)[lb]{\smash{{\SetFigFont{20}{24.0}{\rmdefault}{\mddefault}{\itdefault}{\color[rgb]{0,0,0}$\mathfrak{g}_0$}%
}}}}
\put(2476,-3211){\makebox(0,0)[lb]{\smash{{\SetFigFont{16}{24.0}{\rmdefault}{\mddefault}{\itdefault}{\color[rgb]{0,0,0}algebras}%
}}}}
\put(2476,-2761){\makebox(0,0)[lb]{\smash{{\SetFigFont{16}{24.0}{\rmdefault}{\mddefault}{\itdefault}{\color[rgb]{0,0,0}Comm.}%
}}}}
\put(5176,-3211){\makebox(0,0)[lb]{\smash{{\SetFigFont{16}{24.0}{\rmdefault}{\mddefault}{\itdefault}{\color[rgb]{0,0,0}algebras}%
}}}}
{\color[rgb]{0,0,0}\put(4951,-6361){\line( 1, 0){225}}
}%
\end{picture}%
\end{center}

The left part consists of the class of commutative associative algebras, and the right one of the class of Lie algebras. 
These two classes are known to be Koszul dual. 
The idea is to relate the two classes in a concrete way, using $\Z_2$-graded structures.
A commutative algebra $\mathfrak{a}_0$ is considered as the even part of a  $\Z_2$-commutative
superalgebra $\mathfrak{a}=\mathfrak{a}_0\oplus\mathfrak{a}_1$, and a Lie algebra $\mathfrak{g}_0$ as the even part of a Lie superalgebra $\mathfrak{g}=\mathfrak{g}_0\oplus\mathfrak{g}_1$, each of these even parts being generated by a common odd part $\mathfrak{a}_1=\mathfrak{g}_1$.

\section{Origin and first examples}

\subsection{A new invariant bivector fields}
The origin of Lie antialgebras comes from the following observation.
Consider the supervariety $M=\R^{2,1}$, on which the even coordinates are denoted $p,q$, and the odd coordinate is $\tau$. 
The standard symplectic form on $M$ is 
$$\om=dp\wedge dq+\half d\tau\wedge d\tau.$$
The form $\om$ is preserved by the action of the Lie superalgebra $\osp$. 

A natural question is 
\textit{what are the invariant bivectors fields with respect to the action of $\osp$ ?}

One can immediately exhibit the following invariant bivector, which is the inverse of $\om$,
$$
\Pc=\ddp \wedge \ddq +\half \ddt \wedge \ddt .
$$
It turns out that there is another invariant bivector, given by 
$$
\L=\ddt \wedge \E + \tau\, \ddp \wedge \ddq,
$$
where $\E=p\ddp+q\ddq+\tau \ddt$ is the Euler field.

It seems that the existence of this extra invariant bivector $\L$ had not been observed before Ovsienko's work. 
In addition, Ovsienko showed that there are no other invariant bivectors beside of $\Pc$ and $\L$ and their linear combinations.\\

What are the properties of $\L$? 
One notes that the bivector $\L$ is odd with linear coefficients whereas $\Pc$ is even with constant coefficients. Let us briefly compare the properties of $\L$ and that of $\Pc$.

\subsection{Algebraic structures associated to the bivectors $\mathcal{P}$ and $\Lambda$}

The first thing one can do is to look at algebraic stuctures associated to these two bivectors.
More precisely, given two homogeneous functions $F,G\in C^{\infty}(M)$, one can construct natural products using the duality between bivectors and 2-forms.

$$
\begin{array}{lcrcc}
\{F,G\}&:=&\langle \;\mathcal{P}, dF\wedge dG\; \rangle \quad &\longrightarrow& \quad \text{Lie (Poisson) superalgebra}\\[20pt]
\left]F,G\right[&:=&\frac{-(-1)^{|F|}}{2}\langle \;\Lambda\, , dF\wedge dG\; \rangle \qquad &\longrightarrow& \qquad \text{??}\\[15pt]
\end{array}
$$

The bracket $\{\,,\,\}$ associated to the even bivector $\Pc$ gives a structure of Lie superalgebra on $C^{\infty}(M)$. 
What about the bracket $]\,,\,[$ associated to $\L$ ? 
The sign $(-1)^{p(F)}$, where $|F|$ stands for the parity of $F$, appearing in the definition,
is added 
to make the operation $\left]F,G\right[$ supercommutative (the coefficient of $-\half$ is just a normalization and is not essential).
It is not easy to exhibit general properties of this bracket on the entire space of functions.

\subsection{Examples}
Let us restrict our attention to nice subspaces of $C^{\infty}(M)$.
For instance, the space of quadratic polynomials is stable under the Lie bracket $\{,\}$ and forms a Lie superalgebra isomorphic to $\osp$. 
Similarly, the space of linear functions is stable for the odd bracket $],[$, 
and forms  (after parity inversion) a 3-dimensional algebra which is also well known: it is
isomorphic to the {\it tiny Kaplansky algebra} $K_3$.
Recall that $K_3$ has one even basis vector, say $\e$, and two odds, say $a,b$, subject to the following 
multiplication rules:
$$\textstyle
\e\e=\e, \quad \e a =\half a,\quad \e b =\half b, \quad ab=\half \e
$$
(these relations may be obtained in naming $\e, a,b$ the images of $\tau, q,p$ after parity inversion).\\

More generally, the space of homogeneous functions of degree 2 (here the homogeneity is defined with respect to the Euler field, i.e. $\E(F)=2F$) is stable under $\{\,,\,\}$. 
The subspace of rational functions with poles allowed only at points where $p=0$ or $q=0$, forms a Lie superalgebra isomorphic to the superconformal algebra $\Kc(1)$, also known as 
Neveu-Schwartz  or super Virasoro algebra (see \cite{Ovs} for details).
The parallel situation for the bracket  $]\,,\,[$, is to consider the stable subspace of homogeneous rational functions of degree 1. It forms an infinite-dimensional algebra denoted $AK(1)$ and that can be described as follows: the even basis vectors $\e_n$, $n\in \Z$, and the odd basis vectors $a_i$,  $i\in \Z+\half$, are subject to the multiplication rules
$$\textstyle
\e_n\e_m=\e_{n+m}, \quad \e_n a_i=\half a_{n+i}, \quad a_ia_j=\half(i-j)\e_{i+j},
$$
(these relations maybe obtained in naming $\e_n$ and $a_i$ the images of $ \tau (\frac{q}{p})^{n}$ and $p(\frac{q}{p})^{i+\half}$,  after parity reversion).\\

The situation can be summarized as follows.
$$
\begin{array}{lclcc}
\{ \text{quadratic polynomials}\}&=&<p^2,pq,q^2,p\tau,q\tau> \quad &\overset{\{,\}}{\longrightarrow}& \osp\\[15pt]
\{ \text{linear polynomials}\}&=&<p,q,\tau> &\overset{],[}\longrightarrow& K_3\\[15pt]
\end{array}
$$
$$
\begin{array}{lclcc}
\{ \text{rat. functions of deg 2}\}=<p^2(\frac{q}{p})^{n+1}, \tau p(\frac{q}{p})^{i+\half}, n\in \Z, i\in \Z+\half> &\overset{\{,\}}{\longrightarrow}& \Kc(1)\\[15pt]
\{ \text{rat. functions of deg 1}\}=<p(\frac{q}{p})^{i+\half}, \tau (\frac{q}{p})^{n}, n\in \Z, i\in \Z+\half> &\overset{],[}{\longrightarrow}& AK(1)\\[20pt]
\end{array}
$$

The algebras $K_3$ and $AK(1)$ are both known as Jordan superalgebras.
\begin{ques}\label{q1}
What are the general properties of $]\,,\,[\;$? Can one construct other nice algebras out of this bracket?
Do invariant odd bivector fields analogous to $\Lambda$  exist in higher dimension?
\end{ques}

\section{General theory}

\subsection{Axioms}
The above examples are quite encouraging to try to build out of them a general theory.

\begin{defn}\cite{Ovs}
A Lie antialgebra $\mathfrak{a}=\mathfrak{a}_0\oplus \mathfrak{a}_1$ is a 
supercommutative algebra satisfying
\begin{itemize}
\item[(i)]$\mathfrak{a}_0$ is associative,
\item[(ii)]right multiplications $R_y:\mathfrak{a}\rightarrow \mathfrak{a}$, 
$a\mapsto ay$ are odd derivations for  $y\in \mathfrak{a}_1$,
\item[(iii)]$\mathfrak{a}_0$ acts commutatively on $\mathfrak{a}_1$, i.e. \\
$x_1(x_2y)=x_2(x_1y)$
for $x_1,x_2\in \mathfrak{a}_0,\; y \in \mathfrak{a}_1$.\\
\end{itemize}
\end{defn}

\noindent
Remarks on the axioms:
\begin{itemize}
\item  $\mathfrak{a}_1$ is not a $\mathfrak{a}_0$-module, because of the axiom (ii),  the axiom (iii) is equivalent to 
$$
\text{(iii')} \quad x_1(x_2y)=\half(x_1x_2)y, \; \text{for } \,x_1,x_2\in \mathfrak{a}_0,\; y \in \mathfrak{a}_1,
$$
at a first sight this half-action may appear unnatural but it is actually essential to the theory,
\item the defining identities of Lie antialgebras are all cubic,
\item half-unital  superalgebras 
satisfying (i) and (iii') are called Kaplansky algebras in \cite{McC2}, referring  to constructions of I. Kaplansky \cite{Kap},
\item the axioms (i) and (ii) imply that $\ga$ is a Jordan superalgebra, (it is quite involved to go from these cubic identities to the quartic identites of Jordan superalgebra, one has to find the right sequence of transformations; this was done in details in \cite{McC2} using the weaker requirement that $\ga_0$ is Jordan),
\item if $\ga$ is generated by its odd part, then axioms (ii) and (iii) imply the axiom (i) (this was observed in \cite{LMG}; axiom (iii) plays a crucial role in this property).
\end{itemize}

Lie antialgebras can be understood as a non trivial super analog to commutative associative algebras.

\begin{ex}
The algebras $K_3$ and $AK(1)$ are Lie antialgebras. But, the entire space $C^{\infty}(M)$ equipped with the bracket $]\,,\,[$ is not.
\end{ex}

\subsection{Adjoint Lie superalgebra}
A nice feature of the theory of Lie antialgebras is a relationship to the theory of Lie superalgebras.
The axiomatic definition of Lie antialgebra allows one to construct a formal Lie superalgebra sharing the same odd space.

To any Lie antialgebra $\ga=\ga_0\oplus \ga_1$ associate the Lie superalgebra $\gf(\ga)=\gf_0\oplus\gf_1$, where, as vector spaces, $\gf_1=\ga_1$ and $\gf_0=S^2_{\ga_0}\ga_1$ is the space of symmetric tensors of elements of $\ga_1$ over $\ga_0$. In other words, elements of $\gf_0$ are of the forms $y_1\odot y_2$, where $y_1,y_2\in \ga_1$ and $\odot$ is the symmetric tensor product 
over $\ga_0$, so that
\begin{equation*}\label{relequiv}
\left\{
\begin{array}{rcl}
y_1\odot y_2& = &y_2\odot y_1,\\
y_1x\odot y_2& = &y_1\odot y_2x, \quad \;y_1,y_2\in \ga_1, x \in \ga_0 .\\
\end{array}
\right.
\end{equation*}
Now, define a bracket on $\gf(\ga)$ as follows
\begin{equation}\label{superbracket}
\begin{array}{rcl}
[y_1,y_2]&=& y_1\odot y_2,\\[5pt]
[y_1\odot y_2, y_3]=-[y_3,y_1\odot y_2]&=&y_1(y_2y_3)+y_2(y_1y_3),\\[5pt]
[y_1\odot y_2, y_3\odot y_4]&=&[y_1\odot y_2, y_3] \odot y_4 + [y_1\odot y_2, y_4]\odot y_3,
\end{array}
\end{equation}
where $y_1,y_2,y_3$ and $y_4$ are elements of $\gf_1=\ga_1$.

The above bracket is the most natural skew-symmetric bracket one may think of.
It was announced to be a Lie superbracket in \cite{Ovs} and then the proof was completed in \cite{LMG}.
It is quite amazing how the axioms of Lie antialgebras combine to produce the Lie superalgebra structure. Once again here the axiom (iii) plays a crucial role. 
To convince the reader, we suggest as an exercice to show that the bracket on $\gf(\ga)$ is well defined, i.e. to show that $[y_1\odot  y_2x, y_3]=[y_1x\odot y_2, y_3].$

\begin{rem} Let us stress that the above construction is different from the well known Tits-Kantor-Koecher (TKK) construction.
The adjoint Lie superalgebras associated to $K_3$ and $AK(1)$ are
$$
\gf(K_3)=\osp,\quad \gf(AK(1))=\Kc(1).
$$
In these examples, the adjoint algebras coincide with the algebra of derivations. 
But, in general this is not true. One has only an embedding
$$
\gf(\ga)\hookrightarrow \Der(\ga).$$
\end{rem}

The following problems remain open.

\begin{ques}
Is there a natural construction of a Lie antialgebra from a Lie superalgebra? 
Is there a direct link between the constructed adjoint Lie superalgebra $\gf(\mathfrak{a})$ and the one obtained through the TKK process?
\end{ques}

\begin{ques}
The relation $\osp =\Der(K_3)$ is crucial in the construction of \cite{BeEl}. Can we expect similar construction from $\Kc(1)=\Der(AK(1))$, or more generally from $\ga$ and $\gf(\ga)$?
\end{ques}

\subsection{Representation theory}
Lie antialgebras are connected to three large classes of algebras: commutative associative, 
Jordan, and Lie.
As a particular class of Jordan algebras, one may apply the 
classical representation theory of Jordan algebras to Lie antialgebras, but
 it seems important to adapt a little bit the classical definitions to take into account the specificity of Lie antialgebras.

\begin{defn}\cite{Ovs},\cite{LMG}
A linear map $\varrho: \mathfrak{a}\rightarrow \mathrm{End}(V)$, where $V$ is a $\Z_2$-graded space, is a representation of Lie antialgebra  if 
\begin{enumerate}
\item $\varrho(ab)=[\varrho(a)\, ,\, \varrho(b)]_+$, for all $a,b \in \mathfrak{a}$,and
\item $\varrho(x_1x_2)=\varrho(x_1)\varrho(x_2)$, for all $x_1,x_2\in \mathfrak{a}_0$.
\end{enumerate}
\end{defn}
The notation $[\,,\,]_+$ stands for the usual Jordan superbracket constructed out of an associative superalgebra, i.e. on homogeneous elements
$$
[X,Y]_+=\half(XY+(-1)^{|X||Y|}YX),
$$
whereas the ususal commutator will be denoted $[\,,\,]$
$$
[X,Y]=XY-(-1)^{|X||Y|}YX.
$$
In other words, a representation of Lie antialgebra is a Jordan representation with the extra requirement that the restriction to the even part is a morphism of associative algebra.\\

A nice feature of the theory is the following relationship between representations of $\ga$ and those of $\gf(\ga)$.

\begin{Thm}\cite{LMG}
Consider a Lie antialgebra $\ga$ and its adjoint Lie superalgebra $\gf(\ga)$.
\begin{enumerate}
\item Every representation $\varrho: \ga \rightarrow \mathrm{End}(V)$ induces a unique representation 
$\tilde{\varrho}: \gf(\ga) \rightarrow \mathrm{End}(V)$ such that $\tilde{\varrho}(\gf_1)=\varrho(\ga_1)$.
\item There exists an ideal $\Ic_{\ga}$ in the universal enveloping algebra $U(\gf(\ga))$ such that every representation $\tilde{\varrho}: \gf(\ga) \rightarrow \mathrm{End}(V)$ vanishing on $\Ic_{\ga}$ induces a unique representation $\varrho: \ga \rightarrow \mathrm{End}(V)$ with $\tilde{\varrho}(\gf_1)=\varrho(\ga_1)$.
\end{enumerate}
\end{Thm}

In other words, point (1) of the above theorem says that the images of odd elements of $\ga$ 
generate  in $\left(\mathrm{End}(V), [\,,\,]\right)$
a representation of $\gf(\ga)$. 
This is quite surprising. 

Consider for example an arbitrary representation of $K_3$, and denote by $A,B, \E$ the images of $a,b,\e$. The following relations hold
\begin{equation*}
\left \lbrace
\begin{array}{rcl}
AB -BA&=&\E\\ [5pt] 
A\E + \E A&=& A\\ [5pt]
B\E + \E B&=& B\\ [5pt]
 \E^2&=&\E.
\end{array}
\right.
\end{equation*}
Now, denote by $H:=-(AB+BA)$, $E=A^2$ and $F:=-B^2$. These three elements automatically satisfy the relations of 
$\mathrm{ \bf sl}(2)$, i.e.
\begin{equation*}
\left \lbrace
\begin{array}{rcl}
HE-EH&=&2E\\ [5pt] 
HF-FH&=& -2F\\ [5pt]
EF - FE&=& H,
\end{array}
\right.
\end{equation*}
and together with $A,B$ they satisfy the relations of $\osp$.

The converse is not true. The point (2) of the above theorem characterizes the representations of
the Lie superalgebra for which the converse holds. 
In the case of $\ga=K_3$, the ideal $\Ic_\ga$ of $U(\osp)$ is the ideal generated by the Casimir element.

More details on the representation of $K_3$ viewed as a Lie antialgebra can be found in \cite{MG}.
Elements of general representation theory for Lie antialgebras are developed in \cite{LMG}.

\begin{ques}
In \cite{LMG} the notion of universal enveloping algebra for a Lie antialgebra is studied and it is observed that $U(K_3)$ appears in a form similar to the generalized Weyl algebras (2.1)  in \cite{BaOy}.
Are there more relations between  Lie antialgebras and generalized Weyl algebras?
\end{ques}

\section{More examples coming from geometry}

In this last section, we describe a series of Lie antialgebras closely related to the Lie algebras of Krichever-Novikov. This series generalized $AK(1)$ in the same way as Krichever-Novikov algebras generalize the Witt algebra.

Consider a Riemann surface $\Sigma$ of genus $g$ with a set $\M$ of $N$ marked points.
The marked points are the points where poles of meromorphic functions on $\Sigma$ are allowed.

Denote by $\Fc_\lambda$ the family of modules of meromorphic tensor densities of weight $\lambda$.
In order to well define these modules for any complex parameter $\lambda$, one should fix a complex logarithm on $\Sigma$.  In what follows we will only consider the cases where $\lambda$ is integer or half integer. 

The space $\Fc_0$ is well identified, it is simply the set of meromorphic functions on $\Sigma$, holomorphic outside of $\M$. The space $\Fc_{-1}$ is also well identified, it is the Lie algebra of meromorphic vector fields on $\Sigma$ holomorphic outside of $\M$. 

In the case of genus $g=0$ and $N=2$ marked points, the Lie algebra $\Fc_{-1}$
is the famous Witt algebra. In higher genus $g\geq 0$ and $N=2$ marked points, the Lie algebras $\Fc_{-1}$
are known as Krichever-Novikov algebras, see \cite{KN1}, \cite{KN2}, \cite{KN3}. 
The general case, $g\geq0$ and $N\geq 2$, has been extensively studied by Schlichenmaier, \cite{Schli2}, \cite{Schli3}, \cite{Schli}, see also \cite{Dic}.

One has natural actions of $\Fc_{-1}$ on itself (by commutator) and on $\Fc_0$ (by derivation).
These actions can be unified and naturally deformed in a one-parameter family of actions, giving rise to the modules $\Fc_\lambda$.
Using local coordinates, the actions of $\Fc_{-1}$ can be described as follows:

$$
\begin{array}{lcll}
&\Fc_{-1}\times \Fc_{\lambda} &\longrightarrow &\Fc_{\lambda}\\[8pt]
&\big(f(z)dz^{-1},g(z)dz^{\lambda}\big)&\mapsto &\big( f(z)g'(z)+\lambda f'(z)g(z)\big)dz^{\l}.
\end{array}
$$
This operation can be generalized in a skew-symmetric operation between two density modules
$$
\begin{array}{lcll}
\{ \;,\;\} : &\Fc_{\mu}\times \Fc_{\l} &\longrightarrow &\Fc_{\l+\mu+1}\\[8pt]
&\big(f(z)dz^{\mu},g(z)dz^{\l}\big)&\mapsto &\big(-\mu f(z)g'(z)+\l f'(z)g(z)\big)dz^{\l+\mu+1}.
\end{array}
$$
Together, with the natural multiplication of two tensor densities
$$
\begin{array}{lcll}
\bullet \;: &\Fc_{\mu}\times \Fc_{\l} &\longrightarrow &\Fc_{\l+\mu}\\[8pt]
&\big(f(z)dz^{\mu},g(z)dz^{\l}\big)&\mapsto &f(z)g(z)dz^{\l+\mu}
\end{array}
$$
they provide a natural structure of Poisson algebra on the graded space of all tensor densities
$\bigoplus_\l \Fc_\l$.

\begin{Thm}\cite{LMG2}
The $\Z_2$-graded space $\Fc_0\oplus \Fc_{-\half}$ has a natural structure of a Lie antialgebra, given by the product
\begin{equation*}\label{JKN}
\begin{array}{rcl}
f(z)\circ g(z)&=&f(z)\bullet  g(z)\\[8pt]
f(z)\circ \g(z)(dz)^{-\half}&=&\half\,f(z)\bullet \g(z)(dz)^{-\half}\\[8pt]
\vp(z)(dz)^{-\half}\circ \g(z)(dz)^{-\half}&=&\left\{\vp(z)(dz)^{-\half}, \g(z)(dz)^{-\half}\right\}.
\end{array}
\end{equation*} 
Its adjoint Lie superalgebra is the space $\Fc_{-1}\oplus \Fc_{-\half}$, equipped with the bracket
\begin{equation*}\label{LKN}
\begin{array}{rcl}
\left[f(z)(dz)^{-1}\;,\;g(z)(dz)^{-1}\right]&=&\left\{ f(z)(dz)^{-1}, g(z)(dz)^{-1}\right\}\\[8pt]
\left[f(z)(dz)^{-1}\;,\;\g(z)(dz)^{-\half}\right]&=&\left\{f(z)(dz)^{-1}, \g(z)(dz)^{-\half}\right\}\\[8pt]
\left[\vp(z)(dz)^{-\half}\;,\;\g(z)(dz)^{-\half}\right]&=&\half \,\vp(z)(dz)^{-\half}\bullet \g(z)(dz)^{-\half}.\\
\end{array}
\end{equation*}
\end{Thm}

In the case $g=0$, $N=2$, the Lie antialgebra is nothing but $AK(1)$. In the case $g=0$, $N=3$, the Lie antialgebra is closely related to a Jordan superalgebra algebraically constructed in \cite{Zhe1} to provide a ``new type'' of Jordan superalgebra. 
Let us also mention computations of cocycles on the Lie antialgebras and Lie superalgebras of Krichever-Novikov type \cite{Kre}. \\

This construction gives a concrete realization of our initial picture!\\

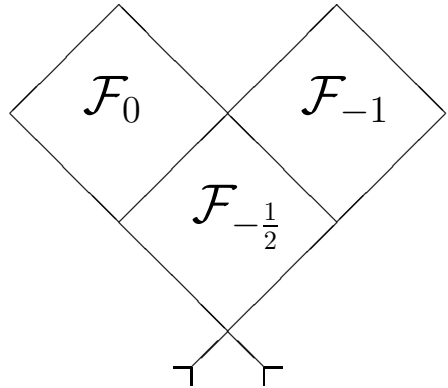
\setlength{\unitlength}{2000sp}%
\begingroup\makeatletter\ifx\SetFigFont\undefined%
\gdef\SetFigFont#1#2#3#4#5{%
  \reset@font\fontsize{#1}{#2pt}%
  \fontfamily{#3}\fontseries{#4}\fontshape{#5}%
  \selectfont}%
\fi\endgroup%
\begin{center}
\begin{picture}(5424,4749)(1789,-6598)
\put(4051,-4561){\makebox(0,0)[lb]{\smash{{\SetFigFont{20}{24.0}{\rmdefault}{\mddefault}{\itdefault}{\color[rgb]{0,0,0}$\Fc_{-\half}$}%
}}}}
\thinlines
{\color[rgb]{0,0,0}\put(4501,-3211){\line( 1, 1){1350}}
\put(5851,-1861){\line( 1,-1){1350}}
\put(7201,-3211){\line(-1,-1){1350}}
\put(5851,-4561){\line(-1, 1){1350}}
}%
{\color[rgb]{0,0,0}\put(3151,-4561){\line( 1,-1){1350}}
\put(4501,-5911){\line( 1, 1){1350}}
}%
{\color[rgb]{0,0,0}\put(4501,-5911){\line(-1,-1){450}}
}%
{\color[rgb]{0,0,0}\put(4501,-5911){\line( 1,-1){450}}
}%
{\color[rgb]{0,0,0}\put(4051,-6361){\line(-1, 0){225}}
}%
{\color[rgb]{0,0,0}\put(4051,-6361){\line(-1,-1){225}}
}%
{\color[rgb]{0,0,0}\put(4051,-6361){\line( 0,-1){225}}
}%
{\color[rgb]{0,0,0}\put(4951,-6361){\line( 0,-1){225}}
}%
{\color[rgb]{0,0,0}\put(4951,-6361){\line( 1,-1){225}}
}%
{\color[rgb]{0,0,0}\put(4951,-6361){\line( 1, 0){225}}
}%
\put(5401,-3211){\makebox(0,0)[lb]{\smash{{\SetFigFont{20}{24.0}{\rmdefault}{\mddefault}{\itdefault}{\color[rgb]{0,0,0}$\Fc_{-1}$}%
}}}}
\put(2701,-3211){\makebox(0,0)[lb]{\smash{{\SetFigFont{20}{24.0}{\rmdefault}{\mddefault}{\itdefault}{\color[rgb]{0,0,0}$\Fc_0$}%
}}}}
{\color[rgb]{0,0,0}\put(3151,-1861){\line(-1,-1){1350}}
\put(1801,-3211){\line( 1,-1){1350}}
\put(3151,-4561){\line( 1, 1){1350}}
\put(4501,-3211){\line(-1, 1){1350}}
}%
\end{picture}%
\end{center}

\begin{ques}\label{qder}
Krichever-Novikov algebras carry a structure of almost-graded algebras, see \cite{Schli3}. 
How Lie antialgebras behave with respect to this almost-grading?
\end{ques}
\section{Conclusion}

Lie antialgebras are friendly hybrid birds, suggesting a non-trivial super analog to commutative associative algebras, and offering more relations between the theory of Jordan algebras and Lie algebras.
Many directions of the theory need to be explored. 
Questions \ref{q1}-\ref{qder} addressed throughout the paper are examples of precise questions naturally arising in the theory. 
More general open questions, as the use of Lie antialgebras in mathematical physics or link to integrable systems, 
are natural directions of investigations.\\

\noindent
\section*{Acknowledgements} The authors are grateful to V. Ovsienko and M. Schlichenmaier for helpful remarks, references and suggestions.

\section*{References}


\end{document}